
\parindent 20pt
\parskip = 3pt
\def\up#1{$\vphantom x$\vskip -#1pt}
\magnification = \magstep1
\baselineskip= 14pt

\newcount\ftnumber
\def\ft#1{\global\advance\ftnumber by 1
          {\baselineskip=13pt \footnote{$^{\the\ftnumber}$}{#1 }}}

\input colordvi

\font\title = cmr10 scaled 1440        


\def\>{\rangle}
\def\<{\langle}

\def\fr#1/#2{{\textstyle{#1\over#2}}}

\input soul.sty 
\up{16.5}


\vskip 5pt

\centerline{\title  Why QBism is not the Copenhagen interpretation}

\vskip 8pt
\centerline {\title and what John Bell might have thought of it}
\vskip 15pt

{\narrower\narrower
\noindent Based on a talk\ft{Video available at: https://phaidra.univie.ac.at/o:360625}
at the conference ``Quantum [Un]Speak\-ables II: 50 Years of Bell's Theorem'', University of Vienna, June 19, 2014.\ft{To be published in 
{\it Quantum [Un]Speakables II\/} (Springer Verlag, 2015).} 

}

\vskip 25pt
\centerline{ N. David Mermin}

\vskip 10pt

\centerline{Laboratory of Atomic and Solid State Physics}

\centerline{Cornell University, Ithaca NY 14853-2501}
\vskip 50pt
{\narrower
{\it Abstract.\/} Christopher Fuchs and R\"udiger Schack have developed a way of understanding science, which, among other things, resolves many of the conceptual puzzles of quantum mechanics that have vexed people for the past nine decades.  They call it QBism.   I  speculate on how John Bell might have reacted to QBism, and I explain the many ways in which QBism differs importantly from the orthodox ways of thinking about quantum mechanics associated with the term ``Copenhagen interpretation.''    

}
\vfil\eject

\up{-100}

{\narrower\narrower\narrower

\noindent {\sl  Dedicated to  my friend and Cornell colleague Geoffrey Chester (1928-2014), who for over 50 years enjoyed my more controversial enthusiasms, while always insisting that I keep my feet firmly on the ground.}

}

\vfil\eject

\centerline{\bf  How QBism differs from the Copenhagen interpretation}
\vskip 2pt
\centerline {\bf and what John Bell might have thought of it}
\vskip 25pt

{\narrower\narrower
 
\noindent{\sl Our students learn quantum mechanics 
the way they learn to ride bicycles 
(both very valuable accomplishments) 
without really knowing what they are doing.  }

\hskip 30pt  ---J.~S.~Bell, letter  to R.~E.~Peierls, 20/8/1980

\vskip 20 pt
 
 
 
\noindent {\sl I think we invent concepts, like ``particle'' or ``Professor Peierls'', 
 to make the immediate sense of data more intelligible. }

 \hskip 30pt  ---J.~S.~Bell, letter to R.~E.~Peierls, 24/2/1983 
\vskip 20pt
\noindent {\sl I have the impression as I write this,  that a moment ago
I heard the bell of the tea trolley.   But I am not sure
 because  I was concentrating  on what I was writing$\ldots$
  The ideal instantaneous measurements of the textbooks 
  are not precisely realized anywhere anytime,  and
  more or less realized, more or less all the time,
  more or less everywhere.}
  
\hskip 30pt  ---J.~S.~Bell, letter to R.~E.~Peierls, 28/1/1981\ft{{\it Selected Correspondence of Rudolf Peierls}, v.~2, Sabine Lee [ed], World Sci., 2009.  I have the impression (confirmed at the conference) that all three of these quotations are unfamiliar even to those who, like me, have devoured almost everything John Bell ever wrote about quantum foundations.} 
  \vskip5pt

}

\vskip 20pt

For the past decade and a half Christopher Fuchs and R\"udiger Schack (originally in collaboration with Carlton Caves) have been developing a new way to think about quantum mechanics.   Fuchs and Schack have called it QBism.\ft{C. A. Fuchs and R. Schack, Rev.~Mod.~Phys. {\bf 85}, 1693-1714 (2013).}  Their term originally stood for ``quantum Bayesianism''.   But QBism is  a way of thinking about science quite generally, not just quantum physics,\ft{When the QBist view of science is used to solve classical puzzles  I have suggested calling it CBism;     N.~D.~Mermin, Nature,   {\bf 507}, 421-423, March 27, 2014.}
and it is pertinent even when probabilistic judgments, and therefore ``Bayesianism'',  play no role at all.   
I nevertheless retain the  term QBism, both to acknowledge the history behind it, and because a secondary meaning remains apt in the broader context:  QBism is as big a break with 20th century ways of thinking about  science as Cubism was with 19th century ways of thinking about art.


QBism maintains that my understanding of the world rests entirely on the  experiences that the world has induced in me throughout the course of my life. 
Nothing beyond my personal experience underlies the picture that I have formed of my own external world.\ft{For ``my'', ``me,'' ``I'', you can  read appropriate versions of ``each of us''; the singular personal pronoun is less awkward.   But unadorned ``our'', ``us'', and ``we''  are dangerously ambiguous.  In QBism the first person plural  always means each of us individually;   it never means all of us collectively, unless this is spelled out.    Part of the 90-year confusion at the foundations of quantum mechanics can be attributed to the unacknowledged ambiguity of the first-person plural pronouns and the carelessness with which they are almost always used.} 
This is a statement of empiricism.    But it is empiricism taken more seriously than most scientists are willing to do.  

To state that my understanding of the world rests on my experience is not to say that my world exists only within my head, as recent popularizations of QBism have wrongly asserted.\ft{H.~C.~ von Baeyer,  Scientific American  {\bf 308}, 46-51, June 2013; M.~Chalmers, New Scientist, 32-35, May 10, 2014.  I believe that in both cases these gross distortions were the fault of overly intrusive copy editors and headline writers, who did not understand the manuscripts they were trying to improve.}   Among the ingredients from which I construct my picture of my external world are the impact of that world on my experience, when it responds to the actions that I take on it.  When I act on my world,  I generally have no control over how it acts back on me.

Nor does QBism maintain that  each of us is free to construct our own private worlds.   Facile charges of solipsism miss the point.    My experience of you leads me to hypothesize that you are a being very much like myself, with your own private experience. This is
as firm a belief as any I have. I could not function without it. If asked to assign this
hypothesis a probability I would choose $p=1$.\ft{I have more to say about $p=1$ below.}   Although I have no direct personal access to your own experience,   an important component of my private experience is the impact on me of your efforts to communicate, in speech or writing, your verbal representations of your own experience.    Science is a collaborative human effort to find, through our individual actions on the world and our verbal communications with each other, a model for what is common to all of our privately constructed external worlds.   Conversations, conferences, research papers, and books are an essential part of the scientific process.

Fuchs himself may be  partly responsible for the silly accusations about solipsism.   One of his favorite slogans about QBism is\ft{Christopher A.~Fuchs, arXiv:1003.5182.}
``Quantum mechanics is a single-user theory", sometimes abbreviated to\ft{Christopher A. Fuchs, arXiv:1405.2390, especially pp.~546-549.} ``Me, me, me!''   This invites the $s$-word.  I hurled it at him myself the first time I came upon such slogans.  Although susceptible to misinterpretation, they are important reminders that any application of quantum mechanics must ultimately be understood to be undertaken by a particular person\ft{Generally named Alice.}  to help her make sense of her own particular experience.     They were never intended to mean that there cannot be many different users of quantum mechanics.    Nor do they require any particular user to exclude from her own experience what she has heard or read about the private experience of others. 

Those who reject QBism --- currently a large majority of the physicists who know anything about it ---  reify the common external world we have all negotiated with each other, removing  from the story any reference to the origins of our common world in the private experiences we try to share with each other through language.
For all practical purposes reification is a sound strategy.  It would be hard to live our daily private or professional scientific lives if we insisted on constantly tracing every aspect of our external world back to its sources in our own private personal experience.   My reification of the concepts I invent, to make my immediate sense of data more intelligible, is a useful tool of day-to-day living.  

But when subtle conceptual issues are at stake, related to certain notoriously murky scientific  concepts like quantum states,  then we can no longer refuse to acknowledge that our scientific pictures of the world rest on the private experiences of individual scientists.   The most famous investigator Vienna has ever produced, who worked just a short walk from the lecture hall for this conference, put it concisely:   ``A world constitution that takes no account 
of the mental apparatus by which we perceive it is an empty abstraction."     This was said not by Ludwig Boltzmann, not by Erwin Schr\"odinger, and not  even by Anton Zeilinger.   It was said by Sigmund Freud,\ft{{\it The Future of an Illusion}, 1927, concluding paragraph.}  just down the hill at Berggasse 19.  He was writing about religion, but his remark applies equally well to science. 

 After he returned to Vienna in the early 1960s, Schr\"odinger repeatedly made much the same point,\ft{{\it Nature and the Greeks,  Science and Humanism,}  
 Cambridge (1996), p. 92.
 See also {\it Mind and Matter\/} and 
{\it My View of the World\/}.} somewhat less concisely than Freud: ``The scientist subconsciously, almost inadvertently simplifies his problem of understanding Nature by disregarding or cutting out of the picture to be constructed,
himself, his own personality, the subject of cognizance.''  In expressing these views in the 1960s he rarely mentions quantum mechanics.   Only thirty years earlier, in a letter to Sommerfeld, does he explicitly tie this view to quantum mechanics, and even then, he allows that it applies to science much more broadly:  ``Quantum mechanics forbids statements about what really
exists --- statements about the object.  It deals only with
the object-subject relation.  Even though this holds,
after all, for any description of nature, it evidently holds 
in quantum mechanics in a much more radical sense."\ft{Schr\"odinger to Sommerfeld, 11 December, 1931, in {\it Sch\"odingers Briefwechsel zur Wel\-lenmechanik und zum Katzenparadoxon}, Springer Verlag, 2011.}  We were rather successful excluding the subject from classical physics (but not completely$^5$).   Quantum physics finally forced (or should have forced) us to think harder about the importance of the object-subject relation.

Niels Bohr, whose views on the meaning of quantum mechanics Schr\"odinger rejected, also delivered some remarkably QBist-sounding pronouncements, though by ``experience'' he almost certainly meant the objective readings of large classical instruments and not the personal experience of a particular user of quantum mechanics:  ``In our description of nature the purpose is not
 to disclose the real essence of the phenomena
 but only to track down, so far as it is possible, relations
 between the manifold aspects of our experience."\ft{Niels Bohr, 1929.  In  
 {\it Atomic Theory and the Description of Nature\/}, 
 Cambridge (1934), p.~18.}  Thirty years later he was saying pretty much the same thing: ``Physics is to be regarded not so much as
 the study of something a priori given, 
but as the development of methods for 
ordering and surveying human experience."\ft{Niels Bohr, 1961. In  {\it Essays 1958Ð-1962 on Atomic Physics 
and Human Knowledge\/}, Ox Bow Press, Woodbridge, CT (1987), p. 10.}     Bohr and Schr\"odinger are not the only dissenting pair who might have found some common ground in QBism.

The fact that each of us has a view of our world that rests entirely on our private personal experience has little  bearing on how we actually use our scientific concepts to deal with the world.   But it is central to the philosophical concerns of quantum foundational studies.   Failing to recognize the foundational importance of personal experience creates illusory puzzles or paradoxes.    At their most pernicious, such puzzles motivate unnecessary efforts to reformulate in more complicated ways --- or even to change the observational content of --- theories which have been entirely successful for all practical purposes.

This talk is not addressed to those who take (often without acknowledging it) an idealistic or Platonic position in their philosophical meditations on the nature of quantum mechanics.   They will never be comfortable with QBism.   My talk is intended primarily for the growing minority of philosophically minded physicists who, far from rejecting QBism, are starting to maintain that there is nothing very new in it.\ft{I count this as progress.   The four stages of acceptance of a radical new idea are: (1) It's nonsense; (2) It's well known; (3) It's trivial; (4) I thought of it first.  I'm encouraged to find that stage (2) is now well underway.}  I am thinking of those who maintain that QBism is nothing more than the Copenhagen interpretation.

I may be partly to blame for this misunderstanding.   I have used  the above quotations from Bohr in several  recent essays about QBism, because QBism provides a context in which these quotations finally make unambiguous sense.   While they made sense for Bohr too, it was not  a QBist kind of sense, and I very much doubt that people gave them a QBist reading.    Similarly, my quotation from Freud does not mean that QBism should be identified with psychoanalysis, and the three epigraphs from John Bell at the head of this text should not be taken to mean that I believe QBism had already been put forth by Bell in the early 1980s.   The quotations from Bell's letters to Peierls are only to suggest that John Bell, who strenuously and elegantly identified what is incoherent in Copenhagen,  might not have dismissed QBism as  categorically.
There are many important ways in which QBism is profoundly different from Copenhagen, and from any other way of thinking about quantum mechanics that I know of.    If you are oblivious to these differences, then you have missed the point of QBism.

The primary reason people wrongly identify QBism with Copenhagen is that  QBism, like  most varieties of Copenhagen, takes the quantum state of a system to be not an objective property of that system, but a mathematical tool for thinking about the system.\ft{Heisenberg and Peierls are quite clear about this.   Bohr may well have believed it but never spelled it out as explicitly.  Landau and Lifshitz, on the other hand,  are so determined to eliminate any trace of  humanity from the story that I suspect their flavor of Copenhagen might  reject the view of quantum states as mathematical tools.}    In contrast, in many of the major non-standard interpretations --- many worlds, Bohmian mechanics, and spontaneous collapse theories --- the quantum state of a system is very much an objective property of that system.\ft{In consistent histories, which has a Copenhagen tinge, its quantum state  can be a true property of a system, but only relative to a ``framework''.}    Even people who reject all these heresies and claim to hold standard views of quantum mechanics, are often careless about reifying quantum states.   Some claim, for example, that quantum states were evolving (and even collapsing) in the early universe, long before anybody existed to assign such states.    But the models of the early universe to which we assign quantum states are models that we construct to account for contemporary astrophysical data.   In the absence of such data, we would not have come up with the models.   As Rudolf Peierls remarked, 
``If there is a part of the Universe, or a period in its history, which is not capable of influencing present-day events directly or indirectly, then indeed there would be no sense in applying quantum mechanics to it."\ft{R. E. Peierls, Physics World January 1991, 19-20.}



A fundamental difference between QBism and any flavor of Copenhagen, is that QBism explicitly introduces each {\it user\/} of quantum mechanics into the story, together with the world external to that user.   Since every user is different, dividing the world differently into external and internal, every application of quantum mechanics to the world must ultimately refer, if only implicitly, to a particular user.    But every version of Copenhagen takes a view of the  world that makes no reference to the particular user who is trying to make sense of that world.

Fuchs and Schack prefer the term ``agent'' to ``user''.   ``Agent'' serves to emphasize that the user takes actions on her world and experiences the consequences of her actions.     I prefer the term ``user'' to emphasize Fuchs' and Schack's equally important point that science is a user's manual.   Its purpose is to help each of us  make sense of our private experience induced in us by the world outside of us. 

It is crucial to note from the beginning that ``user'' does not mean a generic body of users.   It means a particular individual person, who is making use of science to bring coherence to her own private perceptions.    I can be a ``user''.   You can be a ``user''.   But we are not jointly a user, because my internal personal experience is inaccessible to you except insofar as I attempt to represent it to you verbally, and vice-versa.   Science is about the interface between the experience of any particular person and the subset of the world that is external to that particular user.\ft{See in this regard my remarks above about the dangers of the first-person plural.}    This is unlike anything in any version of Copenhagen.\ft{And unlike any other way of thinking about quantum mechanics.}    It is central to the QBist understanding of science.     

The notion that science is a tool that each of us can apply to our own private body of personal experience is explicitly renounced by the Landau-Lifshitz version of Copenhagen.  The opening pages of  their {\it Quantum Mechanics\/}\ft{Translated into English by John Bell, who was therefore intimately acquainted with it.}  declare that 
``It must be most decidedly emphasized that we
are here not discussing a process of measurement
in which the physicist-observer takes part.''   They explicitly deny the user any role whatever in the story.
To emphasize this they add ``By measurement, in quantum mechanics, we
understand any process of interaction between
classical and quantum objects, {\it occurring
apart from and independently of any observer.\/}''  [My italics.]
In the second quotation Landau and Lifshitz have, from a QBist point of view, replaced each different member of  the set of possible users by one and the same set of ``classical objects''.   Their insistence on eliminating  human users from the story, both individually and collectively,  leads them to declare that ``It is in principle impossible$\ldots$to formulate the basic concepts of quantum mechanics 
without using classical mechanics.''   Here they make two big mistakes:   they replace the experiences of each user with ``classical mechanics'',  and they confound the diverse experiences of many different users into that single abstract entity.   

Bohr seems not as averse as Landau and Lifshitz\ft{But Peierls identifies their positions, referring to ``the view
of Landau and Lifshitz (and therefore of
Bohr)'' in his {\it Physics World\/} article.  He disagrees with all of them, saying that it is incorrect to require the apparatus to
obey classical physics.}
 to letting scientists into the story, but they come in only as  proprietors of a single large, {\it classical\/} measurement apparatus.     All versions of Copenhagen objectify each of the diverse family of users of science into a single common piece of apparatus.   Doing this obliterates the fundamental  QBist fact that a quantum-mechanical description is always relative to the particular user of quantum mechanics who provides that description.    Replacing that user with an apparatus introduces the notoriously ill-defined ``shifty split'' of the world into quantum and classical, that John Bell so elegantly and correctly deplored.

Bell's split is shifty in two respects.   Its character is not fixed.   It can be the Landau-Lifshitz split between ``classical'' and ``quantum''.   But sometimes it is a split between ``macroscopic and microscopic''.   Or between ``irreversible'' and ``reversible.''  The split is also shifty because its location can freely be moved along the
path between whatever  poles have been used to characterize it.    

There is also a split in QBism, but it is specific to each user.   That it shifts from user to user is the full extent to which the split is ``shifty''.   For any particular user there is nothing shifty about it: the split is between that user's directly perceived internal experience, and the external world that that user infers from her experience.   

Closely related to its systematic suppression of the user,  is the central role in Copenhagen of ``measurement'' and the Copenhagen view of the ``outcome'' of a measurement.    In all versions of Copenhagen a measurement is an interaction between a quantum system and 
 a ``measurement apparatus''.    Depending on the version of Copenhagen, the measurement apparatus could belong to a ``classical" domain beyond the scope of quantum mechanics, or it could itself be given a quantum mechanical description.   But in any version of Copenhagen the {\it outcome\/} of a measurement is  some strictly classical information produced by the measurement apparatus as a number on a digital display, or the position of an ordinary pointer, or a number printed on a piece of paper, or a hole punched somewhere along a long tape --- something like that.   Words like ``macroscopic'' or ``irreversible'' are used at this stage to indicate the objective, substantial, non-quantum character of the outcome of a measurement.    

In QBism, on the other hand, a measurement can be {\it any\/} action taken by {\it any\/} user on her external world.   The outcome of the measurement is the {\it experience\/} the world induces back in that  particular user, through its response to her action.   The QBist view of measurement includes  Copenhagen measurements as a special case, in which the action is carried out with the aid of a measurement apparatus and the user's experience consists of her perceiving the display, the pointer, the marks on the paper, or the hole in the tape produced by that apparatus.      But a QBist ``measurement'' is much broader.   Users are making measurements more or less all the time more or less everywhere.   Every action on her world by every user constitutes a measurement, and  her experience of the world's reaction is its outcome.     Physics is not limited to the outcomes of ``piddling" laboratory tests, as Bell complained about Copenhagen.\ft{John S. Bell,  Physics World  {\bf 3} (8), 33Ð40 (1990).}

In contrast to the Copenhagen interpretation (or any other interpretation I am aware of), in QBism the outcome of a measurement is special to the user taking the action --- a private internal experience of that user.   The user can attempt to communicate that experience verbally to other users, who may hear\ft{As John Bell may have heard the bell of the tea trolley. Hearing something, of course, is a personal  experience.}   her words.    Other users can also observe her action and, under appropriate conditions, experience aspects of the world's reaction closely related to those experienced by the original user.    But in QBism the immediate outcome of a measurement is a private  experience of the person taking the measurement action, quite unlike the public, objective, classical outcome of a Copenhagen\ft{I shall stop adding the phrase ``or any other interpretation'', but in many cases the reader should supply it.} measurement.

Because outcomes of Copenhagen measurements are ``classical'', they are {\it ipso facto\/} real and objective.  Because in QBism an outcome is a personal experience of a user,  it is real only for that user, since that user's immediate experience is private, not directly accessible to any other user.  Because the private measurement outcome of a user is not a part of the experience of any other user, it is not as such real for other users.   Some version of the outcome can enter the experience of other users and become real  for them as well, only if the other users have also experienced aspects of the world's response to the user who took the measurement-action, or if that user has sent them reliable verbal or written reports of her own experience. 

This is, of course, nothing but the famous story of Wigner and his friend, but in QBism Wigner's Friend is transformed from a paradox to a fundamental parable.    Until Wigner manages to share in his friend's experience, it makes sense for him to assign her and her apparatus an entangled state in which her possible reports of her experiences (outcomes)  are strictly correlated with the corresponding pointer readings (digital displays, etc.)~of the apparatus.   

Even versions of Copenhagen that do not prohibit mentioning users, would draw the line at allowing a user to apply quantum mechanics to another user's reports of her own internal experience.   Other users are either ignored entirely (along with {\it the\/} user), or they are implicitly regarded as part of ``the classical world''.  
But in QBism each user may assign quantum states in superposition to all of her still unrealized potential experiences, including possible future communications from users she has yet to hear from.    Asher Peres' famous Copenhagen mantra, ``Unperformed experiments have no results'', becomes the QBist user's tautology: ``Unexperienced experiences are not experienced.''  

Copenhagen, as expounded by Heisenberg and Peierls, holds that quantum states encapsulate ``our knowledge''.   This has a QBist flavor to it.   But it is subject to John Bell's famous objection:  Whose knowledge?   Knowledge about what?\ft{Bell used the word ``information'', not ``knowledge", but his objection has the same force with either term.}   QBism  replaces ``knowledge'' with ``belief".   Unlike ``knowledge'', which implies something underlying it that is known, ``belief'' emphasizes a believer, in this case the user of quantum mechanics.   Bell's questions now have simple answers.   Whose belief does the quantum state encapsulate?   The belief of the person who has made that state assignment.    What is the belief about?   Her belief is about the implications of her past experience for her subsequent experience.      

No version of Copenhagen takes the view that ``knowledge'' is the state of belief of  the particular person who is making use of quantum mechanics to organize her experience.    Peierls may come closest in a little-known 1980 letter to John Bell:\ft{Peierls to Bell, 13/11/1980, {\it Selected Correspondence of Rudolf Peierls\/}, vol. 2, Sabine Lee [ed], World Scientific, 2009, p. 807.} 
``In my view, a description of the laws of physics consists in giving us a set of correlations between successive observations. By observations I mean$\ldots$what our senses can experience. That we have senses and can experience such sensations
is an empirical fact, which has not been deduced
(and in my opinion cannot be deduced) from current physics.''    Had Peierls taken care to  specify that when he said ``we'', ``us'', and ``our'' he meant each of us, acting and responding as a user of quantum mechanics, this would have been an early statement of QBism.   But it seems to me more likely that he was using the first person plural collectively, to mean all of us together, thereby promulgating the Copenhagen confusion that Bell  so vividly condemned.

Copenhagen also comes near  QBism in the emphasis Bohr always placed on the outcomes of measurements being stated in ``ordinary language".  I believe he meant by this that measurement outcomes were necessarily ``classical".   In QBism the outcome of a measurement is the experience the world induces back in the user who acts on the world.   ``Classical" for any user is limited to her experience.\ft{Indeed, the term ``classical'' has no fundamental role to play in the QBist understanding of quantum mechanics.   It can be replaced by ``experience".}   So measurement outcomes in QBism are necessarily classical, in a way that has nothing to do with  language.       Ordinary language comes into the QBist story in a more  crucial way than it comes into the story told by Bohr.     Language is the only means by which different users of quantum mechanics can attempt to compare their own private experiences.   Though I cannot myself experience your own experience, I can experience your verbal attempts to represent to me what you experience.    It is only in this way that we can arrive at a shared understanding of what is common to all our own experiences of our own external worlds.   It is this shared understanding that constitutes the content of science. 

A very important  difference of QBism, not only from Copenhagen, but from virtually all other ways of looking at science, is the meaning of probability 1 (or 0).\ft{A good example to keep in mind is my above mentioned assignment of probability 1 to my belief that you have personal experiences of your own that have for you the same immediate character that my experiences have for me.}      In Copenhagen quantum mechanics, an outcome that has probability 1 is enforced by an objective mechanism.    This was most succinctly put by Einstein, Podolsky and Rosen,\ft{A.~Einstein, B.~Podolsky, and N.~Rosen, Phys.~Rev.~{\bf 47}, 777-780 (1935).} though they were, notoriously,  no fans of Copenhagen.    Probability-1 judgments, they held, were backed up by ``elements of physical reality".  

Bohr\ft{N. Bohr, Phys.~Rev.~{\bf 48}, 696-702 (1935).} held that the mistake of EPR lay in an ``essential ambiguity'' in their phrase ``without in any way disturbing''.   For a QBist, their mistake is much simpler than that:  probability-1 assignments, like more general probability-$p$ assignments are personal expressions of a willingness to place or accept bets, constrained only by the requirement\ft{Known as Dutch-book coherence.  See the Fuchs-Schack Revs.~Mod.~Phys.~article cited above.} that they should not lead to certain loss in any single event.  It is wrong to assert that probability assignments must be backed up by objective facts on the ground, even when $p=1$.      An expectation is assigned probability 1 if it is held as strongly as possible. Probability-1 measures the intensity of a belief: supreme confidence.   It does not imply the existence of a deterministic mechanism.

 We are all used to the fact that with the advent of quantum mechanics, determinism disappeared from physics.    Does it make sense for us to qualify this in a footnote: ``Except when quantum mechanics assigns probability 1 to an outcome"?    Indeed, the point was made over 250 years ago by David Hume in his famous critique of induction.\ft{David Hume, {\it An Enquiry concerning Human Understanding\/} (1748).}   Induction is the principle that if something happens over and over and over again, we can take its occurrence to be a deterministic law of nature.    What basis do we have for believing in induction?    Only that it has worked over and over and over again.
 
 That probability-1 assignments are personal judgments, like any other probability assignments, is essential to the coherence of QBism.   It has the virtue of undermining the temptation to infer any kind of ``nonlocality'' in quantum mechanics from the violation of Bell inequalities.\ft{C.~A.~Fuchs, N.~D.~Mermin, and R.~Schack, Am. J. Phys. {\bf 82}, 749-754 (2014).} Though  it is alien to the normal scientific view of probability,  it is no stranger or unacceptable than Hume's  views of induction.\ft{I would have expected philosophers of science, with an interest in quantum mechanics, to have had some instructive things to say about this connection, but I'm still waiting.}   What is indisputable is that the QBist position on probability 1 bears no relation to any version of Copenhagen.  Even Peierls, who gets closer to QBism than any of the other Copenhagenists, takes probability 1 to be backed up by underlying indisputable objective facts.

\vskip 10pt

Since this is a meeting in celebration of John Bell, I conclude with a few more comments on the quotations from Bell's little-known\ft{I have had no success finding any of them with Google.   For example, there is no point in googling ``Bell bicycle."   `` `John S.~Bell' bicycle" does no better.  Even  `` `John S.~Bell' bicycle quantum" fails to produce anything useful, because there is a brand of bicycle called  ``Quantum",  and Quantum bicycles have bells.} corespondence with  Peierls at the head of my text.  

The first quotation suggests a riddle:  Why is quantum mechanics like a bicycle?     {\it Answer\/}: Because while it  is possible to learn how to use either without knowing what you are doing, it is impossible to make sense of either without taking account of what people actually do with them.

The second quotation indicates Bell's willingness to consider concepts, as fundamental as ``particle'' or the person to whom he is writing his letter, as  ``inventions" that  help him to make better sense of the data that constitute his experience.

The third reveals a willingness to regard  measurements as particular responses of particular people to particular experiences induced in them by their external world.

These are all QBist views.  Does this mean that John Bell was a QBist?   No, of course not --- no more than Niels Bohr or Erwin Schr\"odinger or Rudolf Peierls or Sigmund Freud were QBists.  Nobody before Fuchs and Schack has pursued this point of view to its superficially shocking,\ft{Ninety years after the formulation of quantum mechanics, a resolution of the endless disagreements on the meaning of the theory has to be shocking, to account for why it was not discovered long, long ago.} but logically unavoidable and, ultimately, entirely reasonable conclusions.   
On the other hand, what Bell wrote to Peierls, and the way in which he criticized Copenhagen, lead me to doubt that Bell would have rejected QBism as glibly and superficially as most of his contemporary admirers have done.   

John Bell and Rudolf Peierls are two of my scientific heroes, both for their remarkable, often iconoclastic ideas, and for the exceptional elegance and precision with which they put them forth.   Yet in their earlier correspondence, and in their two short papers in {\it Physics World\/}  at the end of Bell's life, they disagree about almost everything in quantum foundations.  
Peierls disliked the term ``Copenhagen interpretation'' because it wrongly suggested that there were other viable ways of understanding quantum mechanics.  Bell clearly felt that Copenhagen was inadequate and downright incoherent.    I like to think that they too, like Bohr and Schr\"odinger,  might have found common ground in QBism.

\vskip 15pt

\noindent {\it Acknowledgment.\/}   I am grateful to Chris Fuchs and R\"udiger Schack for their patient willingness to continue our arguments about QBism, in  spite  of my inability to get their point for many years.     And I thank them both for their comments on earlier versions of this text.

\bye